\documentclass[12pt]{article}
\usepackage{amssymb}
\usepackage{amscd}

\newtheorem{definition}{Definition}[section]
\newtheorem{theorem}[definition]{Theorem}
\newtheorem{lemma}[definition]{Lemma}
\newtheorem{corollary}[definition]{Corollary}

\newtheorem{example}[definition]{Example}

\newtheorem{problem}[definition]{Problem}
\newtheorem{note}[definition]{Note}

\newtheorem{proposition}[definition]{Proposition}

\def\I{\mathbb I}
\def\N{\mathbb N}

\def\K{\mathbb K}

\def\Z{\mathbb Z}
\def\K{\mathbb K}

\begin{document}
\title{\bf The Tetrahedron algebra,
the Onsager \\ algebra, and the
$\mathfrak{sl}_2$ loop algebra
}
\author{
Brian Hartwig
and 
Paul Terwilliger{\footnote{
Department of Mathematics, University of
Wisconsin, 480 Lincoln Drive, Madison WI 53706-1388 USA
{\tt hartwig@math.wisc.edu,
 terwilli@math.wisc.edu }
}
}}
\date{}

\maketitle
\begin{abstract}
Let $\K$ denote a field with characteristic
$0$ and let $T$ denote an indeterminate.
We give a
presentation for the 
three-point loop algebra
$\mathfrak{sl}_2 \otimes 
 \K\lbrack T, T^{-1},(T-1)^{-1}\rbrack$
 via generators and relations.
This presentation displays $S_4$-symmetry. 
Using this presentation
we obtain a decomposition of
the above loop algebra 
into a direct sum of three subalgebras,
each of which is isomorphic to the Onsager algebra.

 \end{abstract}

\section{Introduction}

\noindent
In a seminal paper by Onsager
\cite{Onsager}
the free energy of the two dimensional
Ising model was computed exactly.
In that paper a certain infinite dimensional
Lie algebra was introduced; this
is now called the Onsager algebra and we will
denote it by $O$.
Over the years $O$
has been investigated
many times in 
connection with solvable lattice models 
\cite{CKOnsn},
\cite{Albert},
\cite{agmpy},
\cite{auyangandperk},
\cite{auyangperk2},
\cite{auyang2},
\cite{Baz},
\cite{McCoyd},
\cite{Dolgra},
\cite{vongehlen},
\cite{Klish1},
\cite{Klish2},
\cite{Klish3},
\cite{Lee},
\cite{uglov1} 
representation theory
\cite{Davfirst},
\cite{Dav},
\cite{vg3}
Kac-Moody algebras
\cite{DateRoan2},
\cite{perk},
\cite{roan}
tridiagonal pairs
\cite{hartwig},
  \cite{TD00},
\cite{LS99},
 \cite{qSerre},
\cite{aw}
and partially orthogonal polynomials
\cite{vg1},
\cite{vg2}.
Let us recall some
results on the mathematical side.
In
\cite{Davfirst},
\cite{Dav}
Davies
classified the irreducible
finite dimensional 
$O$-modules.
In 
\cite{perk}
Perk showed
that
$O$
has a presentation involving
generators $A,B$
and relations
\begin{eqnarray*}
\lbrack A,
\lbrack A,
\lbrack A,
B\rbrack \rbrack \rbrack &=& 
4 \lbrack A,
B\rbrack,
\\
\lbrack B,
\lbrack B,
\lbrack B,
A\rbrack \rbrack \rbrack &=& 
4 \lbrack B,
A\rbrack.
\end{eqnarray*}
In 
\cite{roan}
Roan
obtained an injection
from 
$O$ into the loop algebra 
 $\mathfrak{sl}_2 \otimes \K\lbrack T, T^{-1}\rbrack$
where $\K$ denotes a field with characteristic 0 and
$T$ denotes an indeterminate.
In \cite{DateRoan2}
Date and Roan
used this injection
to link the representation theories
of $O$ and
 $\mathfrak{sl}_2 \otimes \K\lbrack T, T^{-1}\rbrack$.

\medskip
\noindent
In this paper we investigate further 
the relationship
between $O$ and
$\mathfrak{sl}_2$
loop algebras.
But instead of 
working with $\mathfrak{sl}_2 \otimes \K\lbrack T, T^{-1}\rbrack$
we will work with
$\mathfrak{sl}_2 \otimes \K\lbrack T, T^{-1}, (T-1)^{-1}\rbrack$.
This algebra appears in 
\cite{brem3}
and \cite[Section 4.3]{sch};
see also 
\cite{brem1}, 
\cite{brem2}, 
\cite{sch2},
\cite{sch3}.
Our first main result is a presentation 
for 
$\mathfrak{sl}_2 \otimes \K\lbrack T, T^{-1}, (T-1)^{-1}\rbrack$
via generators and relations.
To obtain this
presentation we
define a Lie algebra $\boxtimes$
using generators and relations,
and eventually show that 
$\boxtimes$ is isomorphic
to 
$\mathfrak{sl}_2 \otimes \K\lbrack T, T^{-1}, (T-1)^{-1}\rbrack$.
We remark that our presentation of
$\boxtimes$ displays an $S_4$-symmetry. 
In our second main result 
we use the above presentation
to get a decomposition of $\boxtimes$
into a direct sum of three subalgebras, each of which
is isomorphic to $O$.

\medskip
\noindent 
We now give a formal definition
of
$\boxtimes$, followed by a more detailed
description of our results.

\begin{definition}
\label{def:tet}
\rm
Let $\boxtimes$ denote the Lie algebra over $\K$
that has  generators
\begin{eqnarray}
\label{eq:boxgen}
\lbrace X_{ij} \,|\,i,j\in \I, i\not=j\rbrace
\qquad \qquad \I = \lbrace 0,1,2,3\rbrace
\end{eqnarray}
and the following relations:
\begin{enumerate}
\item For distinct $i,j\in \I$,
\begin{eqnarray*}
X_{ij}+X_{ji} = 0.
\label{eq:rel0}
\end{eqnarray*}
\item For mutually distinct $h,i,j\in \I$,
\begin{eqnarray*}
\lbrack X_{hi},X_{ij}\rbrack = 2X_{hi}+2X_{ij}.
\label{eq:rel1}
\end{eqnarray*}
\item For mutually distinct $h,i,j,k\in \I$,
\begin{eqnarray*}
\lbrack X_{hi},
\lbrack X_{hi},
\lbrack X_{hi},
X_{jk}\rbrack \rbrack \rbrack= 
4 \lbrack X_{hi},
X_{jk}\rbrack.
\label{eq:rel2}
\end{eqnarray*}
\end{enumerate}
We call $\boxtimes$ the {\it Tetrahedron algebra}.
\end{definition}

\noindent
In this paper we will prove:
\begin{itemize}
\item
For mutually distinct $h,i,j
 \in \I$ the elements
$X_{hi}, X_{ij}, X_{jh}$
form a basis for a subalgebra of $\boxtimes$
that is isomorphic to 
$\mathfrak{sl}_2$.
\item
For mutually distinct $h,i,j,k \in \I$
the subalgebra of $\boxtimes$ generated
by $X_{hi}, X_{jk}$ is isomorphic
to $O$.
\item
For distinct $r,s \in \I$ 
the subalgebra
of $\boxtimes$ generated by
\begin{eqnarray*}
\lbrace X_{ij} \,|\,i,j\in \I , i\not=j, (i,j)\not=(r,s), (i,j)\not=(s,r)
\rbrace
\end{eqnarray*}
is isomorphic to 
$\mathfrak{sl}_2 \otimes \K\lbrack T, T^{-1}\rbrack$.
\item
$\boxtimes$ is isomorphic to 
$\mathfrak{sl}_2 \otimes \K\lbrack T, T^{-1} ,(T-1)^{-1} \rbrack$.
\item
 Let $\Omega$ (resp. $\Omega'$) (resp. $\Omega''$) denote the
subalgebra of $\boxtimes$ generated by
$X_{12}, X_{03}$
(resp. 
$X_{23}, X_{01}$)
(resp. 
$X_{31}, X_{02}$). By the second bullet above, each of
$\Omega,\Omega', \Omega''$ is isomorphic to $O$.
Then the $\K$-vector space $\boxtimes$ satisfies
\begin{eqnarray*}
\boxtimes = \Omega + \Omega' + \Omega'' \qquad \qquad (\mbox{direct sum}).
\end{eqnarray*}
\end{itemize}

\section{An $S_4$-action on $\boxtimes$}

In this section we describe how the symmetric
group $S_4$ acts on the Tetrahedron algebra
as a group of automorphisms. We will also
review some notational conventions.

\medskip
\noindent 
We identify $S_4$ with the group of permutations
of $\I$.
We denote elements of $S_4$
using 
the cycle notation. For
example
$(123)$ denotes the element of $S_4$ that 
sends
$1 \mapsto 2 \mapsto 3 \mapsto 1$ and
$0 \mapsto 0$.
The group $S_4$ acts on the set of generators for $\boxtimes$
by permuting the indices. Thus 
each $\tau \in S_4$  sends
\begin{eqnarray}
X_{ij} \quad \mapsto \quad  X_{i^\tau j^\tau}
\qquad \qquad (i,j\in \I, \;i\not=j).
\label{eq:perm}
\end{eqnarray}
This action leaves invariant the defining relations for
$\boxtimes$ and therefore
 induces a 
group homomorphism
$S_4 \rightarrow \mbox{Aut}(\boxtimes)$,
where $\mbox{Aut}(\boxtimes)$ denotes the group of automorphisms
of $\boxtimes$. 
This gives an action of $S_4$ on $\boxtimes $ as a group of automorphisms.
For notational convenience we give certain elements of $S_4$ 
 special names:
\begin{eqnarray}
&&
\prime \,= (123),
\qquad \quad \;\omega \,= (13),
\qquad \qquad d \,= (13)(02),
\label{eq:primelab}
\\
&&
\downarrow \,= (12),
\qquad \qquad \Downarrow \,= (03),
\qquad \qquad *\, = (01)(23).
\label{eq:downarrow}
\end{eqnarray}
The same notation will be used for the images of these
elements in $\mbox{Aut}(\boxtimes)$. For example 
\begin{eqnarray}
\label{def:primemap}
&&X'_{01} = X_{02},
\qquad \qquad
X'_{02} = X_{03},
\qquad \qquad
X'_{03} = X_{01},
\\
\label{def:primemap2}
&&X'_{12} = X_{23},
\qquad \qquad 
X'_{23} = X_{31},
\qquad \qquad 
X'_{31} = X_{12}.
\end{eqnarray}
Throughout this paper all group actions are assumed to
be from the right;
this means that  when we apply a product 
$\tau \sigma$ we apply $\tau$ first and then $\sigma$.
For example
\begin{eqnarray*}
X_{12}^{\prime *}= 
(X_{12}^{\prime})^*=
X_{23}^*=
X_{32}.
\end{eqnarray*}
The following subgroups of $S_4$ will play a role in our discussion.
Observe that $\prime, \omega$ generate a subgroup of
$S_4$ that is isomorphic to the symmetric group $S_3$.
Observe that  $\downarrow, \Downarrow, *$ generate a subgroup of
$S_4$ that is isomorphic to the dihedral group $D_4$.
Observe that $\omega,d$ generate a subgroup of
$S_4$ that is isomorphic to the Klein 4-group $\Z_2 \times \Z_2$.

\begin{note}
\rm
In what follows we will discuss
several Lie algebras and their relationship to $\boxtimes$.
These other Lie algebras possess some automorphisms
that we will denote by
$\prime,
\omega, d,
\downarrow, \Downarrow, *$.
We trust that
for any given automorphism, the algebra on which it
acts will be clear from the context.
\end{note}

\section{The Lie algebra $\mathfrak{sl}_2$}

\noindent In this section we discuss the Lie algebra
$\mathfrak{sl}_2$ and its relationship to
$\boxtimes$.

\begin{definition}
\label{def:sl2v1}
\rm
We let $\mathfrak{sl}_2$ denote the Lie algebra over
$\K$ that 
has a basis $e,f,h$ and Lie bracket
\begin{eqnarray*}
\lbrack h,e\rbrack = 2e,
\qquad
\qquad
\lbrack h,f\rbrack = -2f,
\qquad
\qquad
\lbrack e,f\rbrack = h.
\end{eqnarray*}
\end{definition}

\begin{lemma}
\label{lem:sl2v2}
$\mathfrak{sl}_2$ is isomorphic to the Lie algebra
over $\K$ that 
has basis $X,Y,Z$ and Lie bracket
\begin{eqnarray}
\label{eq:equit}
\lbrack
X,Y
\rbrack =  2X+2Y,
\qquad 
\qquad
\lbrack
Y,Z
\rbrack =  2Y+2Z,
\qquad 
\qquad
\lbrack
Z,X
\rbrack =  2Z+2X.
\end{eqnarray}
An isomorphism with the presentation in Definition
\ref{def:sl2v1} is given by
\begin{eqnarray*}
X \;\rightarrow \; 2e-h,
\qquad \qquad 
Y \;\rightarrow \;-2f-h,
\qquad \qquad 
Z \;\rightarrow \; h.
\end{eqnarray*}
The inverse of this isomorphism is given by
\begin{eqnarray*}
e \;\rightarrow \; (X+Z)/2,
\qquad \qquad 
f \;\rightarrow \;-(Y+Z)/2,
\qquad \qquad 
h \;\rightarrow \;Z.
\end{eqnarray*}
\end{lemma}
\noindent {\it Proof:} 
One readily checks that each map is a homomorphism of
Lie algebras and that the maps are inverses. It follows
that each map is an isomorphism of Lie algebras.
\hfill $\Box $  

\begin{note}
\label{note:sl2}
\rm
For notational convenience, for
the rest of this paper
we identify 
the copy of 
$\mathfrak{sl}_2$
given in
Definition
\ref{def:sl2v1} with the copy given in
Lemma
\ref{lem:sl2v2}, via the isomorphism
given in Lemma
\ref{lem:sl2v2}.
\end{note}

\medskip
\noindent We now describe two automorphisms of
$\mathfrak{sl}_2$ that will play a role later in the paper.

\begin{lemma}
\label{lem:sl2aut}
The following (i), (ii) hold.
\begin{enumerate}
\item
There exists an automorphism
$\prime$ of 
$\mathfrak{sl}_2$
such that 
\begin{eqnarray}
\label{eq:pp}
X'= Y,
\qquad \qquad
Y'= Z,
\qquad \qquad
Z'= X.
\end{eqnarray}
\item
There exists an automorphism $\omega$
of 
$\mathfrak{sl}_2$
such that
\begin{eqnarray*}
X^\omega = -Y, 
\qquad \qquad
Y^\omega = -X, 
\qquad \qquad
Z^\omega = -Z.
\end{eqnarray*}
\end{enumerate}
\end{lemma}
\noindent {\it Proof:} (i)
Define the linear transformation
$\prime :
\mathfrak{sl}_2
\to
\mathfrak{sl}_2$
so that 
(\ref{eq:pp}) holds.
We observe that $\prime$ is a bijection. Using
(\ref{eq:equit}) we find 
$\lbrack u,v\rbrack'=\lbrack u',v'\rbrack$ for
all $u,v \in 
\mathfrak{sl}_2$.
\\
\noindent (ii) Similar to the proof of (i) above.
\hfill $\Box $


\begin{note}
\rm
Referring to 
Lemma
\ref{lem:sl2aut}, the automorphism  
$\prime $ has order 3 and each of $\omega, \prime \,\omega$
has order 2.
Therefore $\prime, \omega$ 
generate a subgroup of
$\mbox{Aut}(
\mathfrak{sl}_2)$ that is isomorphic
to $S_3$.
\end{note}

\begin{proposition}
\label{lem:sl2inj}
Let $h,i,j$ denote mutually distinct elements of $\I$.
Then there exists a unique Lie algebra homomorphism from 
$\mathfrak{sl}_2$ to $\boxtimes$ that sends
\begin{eqnarray*}
X \;\rightarrow \; X_{hi},
\qquad \qquad 
Y \;\rightarrow  \;X_{ij},
\qquad \qquad 
Z \;\rightarrow  \;X_{jh}.
\end{eqnarray*}
\end{proposition}
\noindent {\it Proof:} 
By Definition
\ref{def:tet}(ii)
the elements
$X_{hi}, X_{ij}, X_{jh}$ satisfy
the defining relations 
(\ref{eq:equit}) for 
$\mathfrak{sl}_2$. Therefore the homomorphism exists.
The homomorphism is unique since
$X,Y,Z$ form a basis for
$\mathfrak{sl}_2$. 
\hfill $\Box $  

\begin{note}
\rm
In Section 12 we will show 
that the homomorphism in
Proposition
\ref{lem:sl2inj}
is an injection.
\end{note}

\begin{definition}
\label{def:sthomsl2}
\rm
By the {\it standard homomorphism}
from $\mathfrak{sl}_2$ to $\boxtimes$ 
we mean the homomorphism
in Proposition
\ref{lem:sl2inj} with
$(h,i,j)=(1,2,3)$.
\end{definition}

\noindent We finish this section with a comment.

\begin{lemma}
The following diagrams commute:
\[
\begin{CD}
\mathfrak{sl}_2  @>{\rm {st.}}\;{\rm {hom}}>>    \boxtimes  \\
@V\prime VV                     @VV\prime V \\
\mathfrak{sl}_2  @>{\rm {st.}}\;{\rm hom}>>    \boxtimes 
\end{CD}
\qquad \qquad 
\begin{CD}
\mathfrak{sl}_2  @>{\rm st.}\;{\rm hom}>>    \boxtimes  \\
@V\omega VV                     @VV\omega V \\
\mathfrak{sl}_2  @>{\rm st.}\;{\rm hom}>>    \boxtimes  
\end{CD}
\]
\end{lemma}

\section{The Onsager algebra}

\noindent In this section we discuss the Onsager
algebra and its relationship to the Tetrahedron algebra.

\medskip
\noindent Recall the integers
$\Z = \lbrace 0,\pm 1,\pm 2, \ldots \rbrace $
and the natural numbers $\N= \lbrace 1,2,\ldots \rbrace$.

\begin{definition}
\label{def:ons}
\cite{Onsager}
\rm
Let $O$ denote the Lie algebra over $\K$ with
basis $A_m, G_l,$ $m \in \Z$, $l \in \N$ and Lie bracket
\begin{eqnarray*}
\lbrack A_l, A_m \rbrack &=& 2 G_{l-m}   \qquad \qquad l> m,
\\
\lbrack G_l, A_m \rbrack &=& A_{m+l}-A_{m-l},
\\
\lbrack G_l, G_m \rbrack &=& 0.
\end{eqnarray*}
We call $O$ the {\it Onsager algebra}.
\end{definition}

\begin{note}\rm
In Definition 
\ref{def:ons}
our notation is a bit nonstandard. 
 The elements called $A_m, G_l$ in
 Definition
\ref{def:ons}
correspond to the elements called $A_m/2, G_l/2$
in \cite{DateRoan2}.
We make this adjustment for notational convenience.
\end{note}

\begin{lemma}
\label{lem:onsalt}
\cite{perk}
$O$ is isomorphic to the Lie algebra over
$\K$ that has generators $A,B$ and relations
\begin{eqnarray}
\lbrack A,
\lbrack A,
\lbrack A,
B\rbrack \rbrack \rbrack &=& 
4 \lbrack A,
B\rbrack,
\label{eq:dg1}
\\
\lbrack B,
\lbrack B,
\lbrack B,
A\rbrack \rbrack \rbrack &=& 
4 \lbrack B,
A\rbrack.
\label{eq:dg2}
\end{eqnarray}
An isomorphism with the presentation in Definition
\ref{def:ons} is  given by
\begin{eqnarray*}
A \; \rightarrow \;A_0, \qquad \qquad 
B \;\rightarrow \; A_1.
\end{eqnarray*}
\end{lemma}

\begin{note}
\rm
In what follows we identify
the copy of $O$ given in Definition
\ref{def:ons} with the copy given
Lemma \ref{lem:onsalt}, via the isomorphism
in 
Lemma \ref{lem:onsalt}.
\end{note}

\noindent  We now describe three automorphisms of 
$O$ that will play a role in our discussion.

\begin{lemma}
\label{lem:Odiag}
The following (i)--(iii) hold.
\begin{enumerate}
\item
There exists an automorphism 
$\downarrow$
of
$O$ such that $A^\downarrow= -A$ and $B^\downarrow =B$. 
We have
\begin{eqnarray*}
A_m^\downarrow =(-1)^{m-1}A_m, \qquad \qquad G_l^\downarrow =(-1)^l G_l
\qquad \qquad 
m \in \Z, \;\;l \in \N.
\end{eqnarray*}
\item
There exists an automorphism 
$\Downarrow $
of
$O$ such that $A^\Downarrow =A$ and $B^\Downarrow=-B$.
We have
\begin{eqnarray*}
A_m^\Downarrow =(-1)^m A_m, \qquad \qquad G_l^\Downarrow =(-1)^l G_l
\qquad \qquad 
m \in \Z, \;\;l \in \N.
\end{eqnarray*}
\item
There exists an automorphism $*$ of
$O$ such that $A^*=B$ and $B^*=A$. We have
\begin{eqnarray*}
A_m^*=A_{1-m}, \qquad \qquad G_l^*= -G_l
\qquad \qquad 
m \in \Z, \;\;l \in \N.
\end{eqnarray*}
\end{enumerate}
\end{lemma}
\noindent {\it Proof:} 
In each case the map is invertible 
and respects the defining relations
for $O$.
\hfill $\Box $  

\begin{note}
\rm
The  automorphisms
$\downarrow, \Downarrow, *$
from 
Lemma \ref{lem:Odiag}
generate a subgroup of
$\mbox{Aut}(O)$ that is isomorphic to
$D_4$.
\end{note}

\begin{proposition}
\label{prop:onshom}
Let $h,i,j,k$ denote mutually distinct elements of $\I$.
Then there exists a unique Lie algebra homomorphism from
$O$ to $\boxtimes$ that sends
\begin{eqnarray*}
A \rightarrow X_{hi}, 
\qquad \qquad B \rightarrow X_{jk}.
\end{eqnarray*}
\end{proposition}
\noindent {\it Proof:} 
By Definition
\ref{def:tet}(iii)
the elements $X_{hi}, X_{jk}$
satisfy the defining relations
(\ref{eq:dg1}), (\ref{eq:dg2})
for $O$.
Therefore the homomorphism exists.
The homomorphism is unique since
$A,B$ together generate
$O$.
\hfill $\Box $  

\begin{note}
\rm
In Section 12 we will show that
the homomorphism in Proposition
\ref{prop:onshom} is an injection.
\end{note}

\begin{definition}
\label{def:sthom}
\rm
By the {\it standard homomorphism}
from $O$ to $\boxtimes$ 
we mean the homomorphism
from 
Proposition
\ref{prop:onshom}
with 
$(h,i,j,k)=(1,2,0,3)$.
\end{definition}

\noindent We finish this section with a comment.

\begin{lemma}
\label{lem:onscom}
The following diagrams commute:
\[
\begin{CD}
O @>{\rm st.}\;{\rm  hom}>>    \boxtimes  \\
@V\downarrow VV                     @VV\downarrow V \\
O  @>{\rm st.}\;{\rm hom}>>    \boxtimes  
\end{CD}
\qquad \qquad
\begin{CD}
O @>{\rm st.}\;{\rm  hom}>>    \boxtimes  \\
@V\Downarrow VV                     @VV\Downarrow V \\
O  @>{\rm st.}\;{\rm hom}>>    \boxtimes  
\end{CD}
\qquad \qquad
\begin{CD}
O @>{\rm st.}\; {\rm hom}>>    \boxtimes  \\
@V*VV                     @VV*V \\
O  @>{\rm st.}\;{\rm hom}>>    \boxtimes  
\end{CD}
\]
\end{lemma}

\section{The 
$\mathfrak{sl}_2$
loop algebra}

\noindent In this section we discuss the
$\mathfrak{sl}_2$ 
loop algebra
and its relationship to
$\boxtimes$.

\begin{definition}
\label{def:loopalg}
\rm
Let $T$ denote an indeterminate. Let 
$\K\lbrack T, T^{-1} \rbrack$ denote the $\K$-algebra consisting
of all Laurant polynomials in $T$ that have coefficients in $\K$.
Let $L(\mathfrak{sl}_2)$ denote the Lie algebra over $\K$
consisting of the $\K$-vector space
$\mathfrak{sl}_2 \otimes \K\lbrack T, T^{-1}\rbrack$
and Lie bracket
\begin{eqnarray*}
\lbrack u\otimes a, v \otimes b\rbrack = \lbrack u,v\rbrack \otimes ab,
\qquad \qquad 
u,v \in \mathfrak{sl}_2,  
\qquad a,b \in  
 \K\lbrack T, T^{-1}\rbrack.
\end{eqnarray*}
We call $L(\mathfrak{sl}_2)$ the 
{\it $\mathfrak{sl}_2$
loop algebra}.
\end{definition}

\noindent
The
$\mathfrak{sl}_2$
loop algebra 
is 
related to the Kac-Moody algebra associated with the
Cartan matrix
\begin{eqnarray*}
A:= 
\left(
\begin{array}{ c c}
2 & -2  \\
-2 & 2 \end{array} 
\right).
\end{eqnarray*}
This is made clear in the following lemma.

\begin{lemma}
\label{lem:loopkac}
\cite[p.~100]{kacmoody}
The loop algebra 
$L(\mathfrak{sl}_2)$ is isomorphic to the Lie algebra over
$\K$ that has generators $e_i, f_i, h_i,$ $i\in \lbrace 0,1\rbrace$
and the following relations:
\begin{eqnarray*}
h_0+h_1 &=& 0,
\\
\lbrack h_i, e_j \rbrack &=& A_{ij}e_j,
\\
\lbrack h_i, f_j \rbrack &=& -A_{ij}f_j,
\\
\lbrack e_i, f_j \rbrack &=& \delta_{ij}h_j,
\\
\lbrack e_i, \lbrack e_i, \lbrack e_i, e_j \rbrack \rbrack\rbrack
&=& 0, \qquad \qquad i \not=j,
\\
\lbrack f_i, \lbrack f_i, \lbrack f_i, f_j \rbrack \rbrack\rbrack
&=& 0, \qquad \qquad i \not=j.
\end{eqnarray*}
An isomorphism is given by
\begin{eqnarray*}
&&e_1 \;\rightarrow \;e\otimes 1,
\qquad \qquad  \;
f_1 \;\rightarrow \; f\otimes 1,
\qquad \qquad \;\; \;\;
h_1 \;\rightarrow \; h\otimes 1,
\\
&&e_0 \;\rightarrow \; f \otimes T,
\qquad \qquad 
f_0 \;\rightarrow \; e\otimes T^{-1}, 
\qquad \qquad 
h_0 \;\rightarrow \;-h \otimes 1.
\end{eqnarray*}
\end{lemma}

\noindent We now give a second presentation for $L(\mathfrak{sl}_2)$.

\begin{lemma}
\label{lem:loop2}
$L(\mathfrak{sl}_2)$ is isomorphic to the Lie algebra over
$\K$ that has generators 
$X_i, Y_i, Z_i,$ $i \in \lbrace 0,1\rbrace$ and the following
relations.
\begin{eqnarray*}
Z_0+Z_1&=&0,
\\
\lbrack X_i,Y_i \rbrack &=& 2X_i+2Y_i,
\\
\lbrack Y_i,Z_i \rbrack &=& 2Y_i+2Z_i,
\\
\lbrack Z_i,X_i \rbrack &=& 2Z_i+2X_i,
\\
\lbrack Y_i, X_j \rbrack &=& 2Y_i+2X_j \;\; \quad \qquad i\not=j,
\\
\lbrack X_i,
\lbrack X_i,
\lbrack X_i,
X_j\rbrack \rbrack \rbrack &=& 
4 \lbrack X_i,
X_j\rbrack,      \qquad \qquad i \not=j,
\label{eq:ydg1}
\\
\lbrack Y_i,
\lbrack Y_i,
\lbrack Y_i,
Y_j\rbrack \rbrack \rbrack &=& 
4 \lbrack Y_i,
Y_j\rbrack,     \qquad \qquad i \not=j.
\label{eq:ydg2}
\end{eqnarray*}
An isomorphism with the presentation in Lemma
\ref{lem:loopkac} is given by
\begin{eqnarray*}
X_i \;\rightarrow \;2e_i-h_i,
\qquad \qquad  
Y_i \;\rightarrow \;-2f_i-h_i,
\qquad \qquad  
Z_i \;\rightarrow \; h_i.
\end{eqnarray*}
The inverse of this isomorphism is given by
\begin{eqnarray*}
e_i \;\rightarrow \;(X_i+Z_i)/2,
\qquad \qquad 
f_i \;\rightarrow \;-(Y_i+Z_i)/2,
\qquad \qquad 
h_i \;\rightarrow \;Z_i.
\end{eqnarray*}
\end{lemma}
\noindent {\it Proof:} 
One routinely checks that each map is a homomorphism
of Lie algebras and that the maps are inverses.
It follows that each map is an isomorphism of Lie algebras.
\hfill $\Box $  

\begin{note}
\label{note:ident}
\rm
In what follows we identify the copies of
$L(\mathfrak{sl}_2)$ given in
Definition \ref{def:loopalg},
Lemma \ref{lem:loopkac} and 
Lemma \ref{lem:loop2},
via the
isomorphisms given in
Lemma \ref{lem:loopkac} and 
Lemma \ref{lem:loop2}.
\end{note}

\noindent We now describe two automorphisms
of $L(\mathfrak{sl}_2)$ that will play a role in our discussion.

\begin{lemma}
\label{lem:chevdiag}
The following (i), (ii) hold.
\begin{enumerate}
\item
There exists an automorphism 
$\omega$
of
$L(\mathfrak{sl}_2)$ such that
\begin{eqnarray*}
X_i^\omega=-Y_i, \qquad \qquad 
Y_i^\omega=-X_i, \qquad \qquad 
Z_i^\omega=-Z_i
\qquad \qquad 
i \in \lbrace 0,1\rbrace. 
\end{eqnarray*}
\item
There exists an  automorphism 
$d$
of
$L(\mathfrak{sl}_2)$ 
such that
\begin{eqnarray*}
X_i^d=X_j, \qquad \qquad 
Y_i^d=Y_j, \qquad \qquad 
Z_i^d=Z_j
\qquad \qquad i,j \in \lbrace 0,1\rbrace,\quad i\not=j.
\end{eqnarray*}
\end{enumerate}
\end{lemma}


\begin{note}
\rm
The automorphisms $\omega,d$ from Lemma
\ref{lem:chevdiag}
generate a subgroup of 
$\mbox{Aut}(L(\mathfrak{sl}_2))$ 
that is isomorphic to 
$\Z_2 \times \Z_2$.
\end{note}

\begin{proposition}
\label{prop:loopinj}
Let $h,i,j,k$ denote mutually distinct  elements of $\I$.
Then there exists a unique Lie algebra homomorphism
from $L(\mathfrak{sl}_2)$ to $\boxtimes$ that sends
\begin{eqnarray*}
&&X_1\;\rightarrow X_{hi},
\qquad \qquad 
Y_1\;\rightarrow X_{ij},
\qquad \qquad 
Z_1\;\rightarrow X_{jh},
\\
&&X_0\;\rightarrow X_{jk},
\qquad \qquad 
Y_0\;\rightarrow X_{kh},
\qquad \qquad 
Z_0\;\rightarrow X_{hj}.
\end{eqnarray*}
\end{proposition}
\noindent {\it Proof:} 
Comparing the relations
given in Lemma
\ref{lem:loop2}
with the  relations given in
Definition
\ref{def:tet}, we find the homomorphism exists.
This homomorphism is
unique since $X_i, Y_i, Z_i,$ $i\in\lbrace 0,1\rbrace$
is a generating set for 
$L(\mathfrak{sl}_2)$.
\hfill $\Box $  

\begin{note}
\rm In Section 12  we will show 
 that
the homomorphism in Proposition
\ref{prop:loopinj}
is an injection.
\end{note}

\begin{definition}
\rm
By the {\it standard homomorphism}
from $L(\mathfrak{sl}_2)$ to $\boxtimes$ 
we mean the homomorphism
from Proposition
\ref{prop:loopinj} 
with $(h,i,j,k)=(1,2,3,0)$.
\end{definition}

\noindent We finish this section with a comment.

\begin{lemma}
The following diagrams commute:
\[
\begin{CD}
L(\mathfrak{sl}_2)  @>{\rm st.}\;{\rm  hom}>>    \boxtimes  \\
@V\omega VV                     @VV\omega V \\
L(\mathfrak{sl}_2)  @>{\rm st.}\; {\rm hom}>>    \boxtimes  
\end{CD}
\qquad \qquad 
\begin{CD}
L(\mathfrak{sl}_2)  @>{\rm st.}\;{\rm hom}>>    \boxtimes  \\
@VdVV                     @VVdV \\
L(\mathfrak{sl}_2)  @>{\rm st.}\;{\rm hom}>>    \boxtimes  
\end{CD}
\]
\end{lemma}

\section{The three-point 
$\mathfrak{sl}_2$
loop algebra 
}

\noindent In this section we consider an extension of
 $L(\mathfrak{sl}_2)$ that we call
 $L(\mathfrak{sl}_2)^+$.
This algebra is defined as follows.

\begin{definition}
\label{def:extsl2}
\rm
We abbreviate  $\mathcal A$ for the $\K$-algebra 
$\K\lbrack T, T^{-1},(T-1)^{-1}\rbrack$,
where $T$ is indeterminate.
Let $L(\mathfrak{sl}_2)^+$ denote
 the Lie algebra over $\K$
consisting of the $\K$-vector space
$\mathfrak{sl}_2 \otimes \mathcal A
$
and Lie bracket
\begin{eqnarray}
\lbrack u\otimes a, v \otimes b\rbrack = \lbrack u,v\rbrack \otimes ab,
\qquad \qquad 
u,v \in \mathfrak{sl}_2,  
\qquad a,b \in {\mathcal A}. 
\label{eq:brackdef}
\end{eqnarray}
Following \cite{brem3} we  call $L(\mathfrak{sl}_2)^+$ the 
{\it three-point
$\mathfrak{sl}_2$
loop algebra}.
\end{definition}

\noindent Our next goal
is to display a basis for 
 $L(\mathfrak{sl}_2)^+$.
We start with an observation.

\begin{lemma}
\label{lem:order3}
There exists a unique $\K$-algebra automorphism
$\prime $ 
of $\mathcal A$ such that
$T'=1-T^{-1}$. This automorphism has order 3 and satisfies
\begin{eqnarray}
\label{eq:aaut}
&&T''=(1-T)^{-1},
\qquad \qquad 
T T'=T-1,  
\\
&&T' T''= T'-1,
\qquad \qquad 
 T''T = T''-1.
\label{eq:aaut2}
\end{eqnarray}
\end{lemma}

\begin{lemma}
\label{lem:abasis}
The following is a basis for
the $\K$-vector space $\mathcal A$:
\begin{eqnarray}
\label{eq:abasis}
\lbrace 1 \rbrace
\cup
\lbrace T^i, 
(T')^i,
(T'')^i \;|\;
i \in \N\rbrace.
\end{eqnarray}
\end{lemma}
\noindent {\it Proof:} 
We first claim that the elements
(\ref{eq:abasis}) span 
${\mathcal A}$.
Let ${\mathcal A}_1$ denote the subspace of
$\mathcal A$ spanned by the elements
(\ref{eq:abasis}).
Using the data in
Lemma
\ref{lem:order3} we find
${\mathcal A}_1$ is closed under multiplication
and contains
the generators $T,T^{-1}, (T-1)^{-1}$ of
$\mathcal A$.
Therefore ${\mathcal A}_1 = {\mathcal A}$ and
our claim is proved.
It is routine to check that
the elements
(\ref{eq:abasis}) are linearly independent and hence
form a basis for $\mathcal A$.
\hfill $\Box $  

\begin{lemma}
\label{lem:exbasis}
The following is a basis for the 
$\K$-vector space
$L(\mathfrak{sl}_2)^+$:
\begin{eqnarray*}
\label{eq:exbasis}
&&\lbrace 
X\otimes 1,Y\otimes 1,Z\otimes 1\rbrace \cup
\lbrace 
X\otimes T^i,
Y\otimes T^i,
Z\otimes T^i
\;|\;i \in \N\rbrace
\\
&& \;\cup
\;\;
\lbrace
X\otimes (T')^i,
Y\otimes (T')^i,
Z\otimes (T')^i
\;|\;i \in \N\rbrace
\\
&&\;\cup
\;\;
\lbrace 
X\otimes (T'')^i,
Y\otimes (T'')^i,
Z\otimes (T'')^i
\;|\;i \in \N\rbrace.
\end{eqnarray*}
Here $X,Y,Z$ is the basis for 
$\mathfrak{sl}_2$ given in Lemma
\ref{lem:sl2v2}.
\end{lemma}
\noindent {\it Proof:} 
Combine 
Definition
\ref{def:extsl2} and
Lemma
\ref{lem:abasis}.
\hfill $\Box $ \\  

\noindent We now consider 
how $\boxtimes$ is related to
$L(\mathfrak{sl}_2)^+$.

\begin{proposition}
\label{prop:isostart}
There exists a unique 
Lie algebra homomorphism $ \sigma :
\boxtimes \to
L(\mathfrak{sl}_2)^+$
such that
\begin{eqnarray*}
&&X_{12}^{\sigma}\;=\; X\otimes 1,
\qquad \qquad X_{03}^{\sigma}\;=\;Y\otimes T + Z\otimes (T-1),
\\
&&X_{23}^{\sigma}\;=\;Y\otimes 1,
\qquad \qquad X_{01}^{\sigma}\;=\; Z\otimes T' + X\otimes (T'-1),
\\
&&X_{31}^{\sigma}\;=\;Z\otimes 1,
\qquad \qquad X_{02}^{\sigma}\;=\;X\otimes T'' + Y\otimes (T''-1).
\end{eqnarray*}
Here $X,Y,Z$ is the basis for 
$\mathfrak{sl}_2$ given in Lemma
\ref{lem:sl2v2}.
\end{proposition}
\noindent {\it Proof:} 
Using 
(\ref{eq:equit})
and 
(\ref{eq:brackdef})--(\ref{eq:aaut2})
we find that in the above equations 
the expressions on the right-hand side 
satisfy the defining relations 
for $\boxtimes$ 
given in
Definition
\ref{def:tet}. Therefore the homomorphism exists.
The homomorphism is unique since
$X_{12}, X_{23}, X_{31}, X_{01},X_{02}, X_{03}$
is a generating set for 
$\boxtimes$.
\hfill $\Box $  

\begin{note}
\rm
\noindent
In Section 11 we will show  that the homomorphism
in 
Proposition \ref{prop:isostart}
is an isomorphism.
\end{note}

\noindent We 
now introduce several maps that will be useful
later in the paper.

\begin{lemma}
\label{def:prime2}
\rm
There exists an automorphism
$\prime$ of  $L(\mathfrak{sl}_2)^+$ 
that satisfies 
\begin{eqnarray}
\label{eq:oa}
(u\otimes a)' = u'\otimes a'
\qquad \qquad u \in 
\mathfrak{sl}_2, \qquad a \in {\mathcal A},
\end{eqnarray}
where $u'$ is from
Lemma
\ref{lem:sl2aut}(i)
and
$a'$
is from
Lemma \ref{lem:order3}.
This automorphism
 has order 3.
\end{lemma}
\noindent {\it Proof:} 
Define the linear map $\prime 
: L(\mathfrak{sl}_2)^+ 
\to  L(\mathfrak{sl}_2)^+ $
so that 
(\ref{eq:oa}) holds.
The map has order 3 so it is a bijection.
Using (\ref{eq:brackdef})
we find that the map is 
a homomorphism of Lie algebras.
\hfill $\Box $  \\

\begin{lemma}
\label{lem:comdiag} 
The following diagram commutes:
\[
\begin{CD}
\boxtimes  @>\sigma>>
          L(\mathfrak{sl}_2)^+ \\
@V\prime VV                     @VV\prime V \\
\boxtimes
@>\sigma>> 
          L(\mathfrak{sl}_2)^+
\end{CD}
\]
\end{lemma}
\noindent {\it Proof:} 
Combine the data in
Proposition \ref{prop:isostart}
with 
(\ref{def:primemap}), 
(\ref{def:primemap2}).
\hfill $\Box $  \\

\begin{definition}
\label{def:nat}
\rm
Recall 
$L(\mathfrak{sl}_2)= 
\mathfrak{sl}_2\otimes \K\lbrack T,T^{-1} \rbrack$
by Definition
\ref{def:loopalg}.
Also by Definition
\ref{def:extsl2}
we have 
$L(\mathfrak{sl}_2)^+= 
\mathfrak{sl}_2\otimes {\mathcal A}
$,
where
$\mathcal A=  
 \K\lbrack T,T^{-1}, (T-1)^{-1}\rbrack$.
 The inclusion map
$\K\lbrack T,T^{-1} \rbrack \rightarrow
{\mathcal A}$ 
and the identity  
map on 
$\mathfrak{sl}_2$, together 
induce an injection of Lie algebras
$L(\mathfrak{sl}_2) \rightarrow  
L(\mathfrak{sl}_2)^+$.
We call this the {\it natural homomorphism}.
\end{definition}

\begin{lemma}
\label{lem:comdiagnat}
The following diagram commutes:
\[
\begin{CD}
L(\mathfrak{sl}_2)  @>{\rm {nat.}}\;{\rm {hom}}>>
          L(\mathfrak{sl}_2)^+ \\
@V{\rm st.}\;{\rm hom}VV                     @VV{\rm id}V \\
\boxtimes @>>\sigma>  
          L(\mathfrak{sl}_2)^+ 
\end{CD}
\]
\end{lemma}
\noindent {\it Proof:} 
By Lemma
\ref{lem:loopkac}
and Note \ref{note:ident}
the following 
is a generating set for
$L(\mathfrak{sl}_2)$:
\begin{eqnarray*}
&&e_1=e\otimes 1,
\qquad \qquad  \;
f_1=f\otimes 1,
\qquad \qquad \;\; \;\;
h_1=h\otimes 1,
\\
&&e_0=f \otimes T,
\qquad \qquad 
f_0=e\otimes T^{-1}, 
\qquad \qquad 
h_0=-h \otimes 1.
\end{eqnarray*}
We chase these generators
around the diagram.
We illustrate what happens
for the generator  
$e_0$.
By Lemma
\ref{lem:loop2}
and Note \ref{note:ident} we find
\begin{eqnarray}
\label{eq:e0}
e_0=(X_0+Z_0)/2.
\end{eqnarray}
The standard homomorphism 
from
$L(\mathfrak{sl}_2)$ to $\boxtimes$
is the map in 
Proposition
\ref{prop:loopinj} with $(h,i,j,k)=(1,2,3,0)$.
Applying this map to
(\ref{eq:e0})
we get
\begin{eqnarray}
\label{eq:e0p}
(X_{30}+X_{13})/2.
\end{eqnarray}
We now apply $\sigma$ to
(\ref{eq:e0p}) using
Proposition
\ref{prop:isostart} and
Definition \ref{def:tet}(i); the result is
\begin{eqnarray}
\label{eq:e0pp}
-(Y+Z)\otimes T/2.
\end{eqnarray}
Using Lemma \ref{lem:sl2v2}
and
Note
\ref{note:sl2} we find
(\ref{eq:e0pp}) is equal to
$f\otimes T=e_0$, and this is  also the image
of $e_0$ under the natural homomorphism.
For the other generators the details are similar and ommitted.
\hfill $\Box $

\section{A spanning set for $\boxtimes$}

\noindent In this section we display a spanning
set for $\boxtimes$. Later in
the paper 
it will turn out that
this spanning set is a basis for $\boxtimes$. 

\begin{definition}
\label{def:omega}
\rm
Let $\Omega$ denote the subalgebra of 
$\boxtimes$ generated by $X_{12}, X_{03}$.
We observe that $\Omega$ is the image of
the Onsager algebra 
$O$ under the standard homomorphism 
$O\rightarrow \boxtimes$
from  Definition
\ref{def:sthom}.
\end{definition}

\noindent We have a comment.
\begin{lemma}
\label{lem:omegap}
$\Omega'$ is the subalgebra of $\boxtimes$ generated
by $X_{23}, X_{01}$. $\Omega''$ is the subalgebra
of $\boxtimes$ generated by $X_{31}, X_{02}$.
\end{lemma}

\begin{definition}
\rm
\label{lem:agbasis}
Referring to the standard homomorphism
$O\rightarrow \boxtimes$ from
Definition
\ref{def:sthom},
for $m \in \Z$ we let $a_m$ denote the image of $A_m$.
For $l \in \N$ we let $g_l$ denote the image of $G_l$.
\end{definition}

\begin{lemma}
\label{lem:a0a1}
We have
$a_0 = X_{12}$, 
$a_1 = X_{03}$ and
\begin{eqnarray*}
\label{eq:aa}
\lbrack a_l, a_m \rbrack &=& 2 g_{l-m}   \qquad \qquad l> m,
\\
\label{eq:ga}
\lbrack g_l, a_m \rbrack &=& a_{m+l}-a_{m-l},
\\
\label{eq:gg}
\lbrack g_l, g_m \rbrack &=& 0.
\end{eqnarray*}
\end{lemma}
\noindent {\it Proof:} 
Immediate from 
Definition
\ref{def:ons}
and
Definition
\ref{lem:agbasis}.
\hfill $\Box $  

\begin{lemma}
\label{lem:fourpart}
The following (i)--(iii) hold.
\begin{enumerate}
\item $\Omega$ is spanned by 
\begin{eqnarray}
a_m, \;g_l \qquad \qquad m\in \Z,\qquad  l\in \N.
\label{eq:omegaspan}
\end{eqnarray}
\item $\Omega'$ is spanned by 
\begin{eqnarray}
a_m', \; g_l' \qquad \qquad m\in \Z,\qquad  l\in \N.
\label{eq:omegaprimespan}
\end{eqnarray}
\item $\Omega''$ is spanned by 
\begin{eqnarray}
a_m'', \;g_l'' \qquad \qquad m\in \Z,\qquad  l\in \N.
\label{eq:omegaprimeprimespan}
\end{eqnarray}
\end{enumerate}
\end{lemma}
\noindent {\it Proof:} 
\noindent (i)
Recall $O$ is spanned by 
$A_m, G_l,$ $m \in \Z, l\in \N$.
Applying the standard homomophism
$O\rightarrow \boxtimes$ 
we find
$\Omega$ is spanned by $a_m, g_l$, $m \in \Z, l\in \N$.
\\
\noindent (ii), (iii) Apply the automorphism $\prime$.
\hfill $\Box $   \\

\noindent 
We are going to prove that
the union of
(\ref{eq:omegaspan})--(\ref{eq:omegaprimeprimespan})
is a spanning set for
$\boxtimes$. To do this 
we 
show
that $\boxtimes = \Omega+ \Omega'+\Omega''$.
We will use the following lemma.

\begin{lemma} 
\label{lem:action}
For $m \in \N$,
\begin{eqnarray}
\label{eq:act1}
\lbrack a_0',a_m\rbrack
&=&
-2a_0'+2a_m+4\sum_{i=1}^{m-1} a_i 
-4 \sum_{i=1}^{m-1}g_i,
\\
\label{eq:act2}
\lbrack a_0',a_{1-m}\rbrack
&=&
-2a_0'-2a_{1-m}-4\sum_{i=1}^{m-1}a_{1-i} 
-4 \sum_{i=1}^{m-1}g_i,
\\
\label{eq:act3}
\lbrack a_0',g_m\rbrack
&=&
2\sum_{i=1-m}^{m} a_i
\end{eqnarray}
and
\begin{eqnarray}
\label{eq:act4}
\lbrack a_1',a_m\rbrack
&=&
2a_1'-2a_m-4\sum_{i=1}^{m-1}a_i 
-4 \sum_{i=1}^{m-1} g_i,
\\
\label{eq:act5}
\lbrack a_1',a_{1-m}\rbrack
&=&
 2a_1'+2a_{1-m}+4\sum_{i=1}^{m-1} a_{1-i} 
-4 \sum_{i=1}^{m-1} g_i,
\\
\label{eq:act6}
\lbrack a_1',g_m\rbrack
&=&
2\sum_{i=1-m}^{m} a_i.
\end{eqnarray}
\end{lemma}
\noindent {\it Proof:} 
We first verify 
(\ref{eq:act1})--(\ref{eq:act3}) by induction 
on $m$. We start with the case $m=1$. 
By Lemma
\ref{lem:a0a1} we have
$a_0=X_{12}$ and $a_1=X_{03}$.
Also $X_{12}'=X_{23}$ by 
(\ref{def:primemap2}) so $a_0'=X_{23}$.
We have $\lbrack X_{23}, X_{30} \rbrack = 2X_{23}+2X_{30}$
by Definition
\ref{def:tet}(ii)
and
$X_{30}=-X_{03}$
 by Definition
\ref{def:tet}(i).
Combining  these comments
we find $\lbrack a'_0,a_1 \rbrack = -2a'_0+2a_1$
so 
(\ref{eq:act1}) holds for $m=1$.
We mentioned $a_0=X_{12}$ and $a_0'=X_{23}$.
We have
$\lbrack X_{12}, X_{23} \rbrack = 2X_{12}+2X_{23}$
by Definition
\ref{def:tet}(ii)
and recall 
$\lbrack X_{12}, X_{23} \rbrack =
-
\lbrack X_{23}, X_{12} \rbrack$ by the 
definition of a Lie algebra.
Combining these comments
we find
$\lbrack a'_0, a_0\rbrack = -2a'_0 -2a_0$
so 
(\ref{eq:act2}) holds for $m=1$.
By Lemma
\ref{lem:a0a1}
we find
$
\lbrack a_1, a_0\rbrack
=
2g_1
$.
By the Jacobi identity
\begin{eqnarray*}
\lbrack a'_0, \lbrack a_1, a_0 \rbrack \rbrack
=
\lbrack \lbrack a_0', a_1\rbrack , a_0 \rbrack
+
\lbrack a_1,\lbrack a'_0, a_0 \rbrack \rbrack.
\end{eqnarray*}
In this equation we evaluate the left-hand side
using 
$
\lbrack a_1, a_0\rbrack =
2g_1$
and the right-hand side using
(\ref{eq:act1}), (\ref{eq:act2}) at $m=1$.
We routinely find
$\lbrack a'_0, g_1 \rbrack = 2a_0+2a_1$ so
(\ref{eq:act3}) holds  at $m=1$.
We have now verified
(\ref{eq:act1})--(\ref{eq:act3}) for  $m=1$.
Now for an integer $j\geq 2$ we verify 
(\ref{eq:act1})--(\ref{eq:act3}) for  $m=j$.
By induction we may assume
(\ref{eq:act1})--(\ref{eq:act3}) hold for
 $1 \leq m\leq j-1$.
From Lemma  \ref{lem:a0a1} we find
\begin{eqnarray}
\label{eq:garec}
\lbrack g_1, a_{j-1}\rbrack = a_j-a_{j-2}.
\end{eqnarray}
By the Jacobi identity
\begin{eqnarray}
\label{eq:jacobi3}
\lbrack a_0', \lbrack g_1, a_{j-1}\rbrack \rbrack
=
\lbrack \lbrack a_0',  g_1\rbrack , a_{j-1} \rbrack
+
\lbrack  g_1, \lbrack a_0', a_{j-1}\rbrack \rbrack.
\end{eqnarray}
In 
equation (\ref{eq:jacobi3})
we evaluate the left-hand side
using
(\ref{eq:garec}) and the
lines (\ref{eq:act1}), (\ref{eq:act2}) 
at $1 \leq m \leq j-1$.
Moreover 
we evaluate the right-hand side using
Lemma  \ref{lem:a0a1} and
lines (\ref{eq:act1})--(\ref{eq:act3}) 
at $1 \leq m \leq j-1$. From this
we routinely obtain
(\ref{eq:act1}) at $m=j$.
From Lemma  \ref{lem:a0a1} we find
\begin{eqnarray}
\label{eq:garec2}
\lbrack g_1, a_{2-j}\rbrack =a_{3-j}-a_{1-j}.
\end{eqnarray}
By the Jacobi identity
\begin{eqnarray}
\label{eq:jacobi2}
\lbrack a_0', \lbrack g_1, a_{2-j}\rbrack \rbrack
=
\lbrack \lbrack a_0',  g_1\rbrack , a_{2-j} \rbrack
+
\lbrack  g_1, \lbrack a_0', a_{2-j}\rbrack \rbrack.
\end{eqnarray}
In 
equation (\ref{eq:jacobi2})
we evaluate the left-hand side
using
(\ref{eq:garec2}) and
lines (\ref{eq:act1}), (\ref{eq:act2}) 
at $1 \leq m \leq j-1$.
Moreover 
we evaluate the right-hand side using
Lemma  \ref{lem:a0a1} and
lines (\ref{eq:act1})--(\ref{eq:act3}) 
at $1 \leq m \leq j-1$. From this
we routinely obtain
(\ref{eq:act2}) at $m=j$.
By Lemma
\ref{lem:a0a1}
we find
$\lbrack a_j, a_0\rbrack=2g_j$.
By the Jacobi identity
\begin{eqnarray*}
\lbrack a'_0, \lbrack a_j, a_0 \rbrack \rbrack
=
\lbrack \lbrack a_0', a_j\rbrack , a_0 \rbrack
+
\lbrack a_j,\lbrack a'_0, a_0 \rbrack \rbrack.
\end{eqnarray*}
In this equation we evaluate the left-hand side
using 
$\lbrack a_j, a_0\rbrack=2g_j$
and the right-hand side using
(\ref{eq:act1}) at $m=j$.
From this we routinely obtain 
(\ref{eq:act3}) at $ m=j$.
We have now verified
lines (\ref{eq:act1})--(\ref{eq:act3}).
Next we verify lines 
(\ref{eq:act4})--(\ref{eq:act6}).
To do this we apply the automorphism
$\downarrow \Downarrow$
to (\ref{eq:act1})--(\ref{eq:act3}).
Using 
Lemma \ref{lem:Odiag}(i),(ii)
and 
Lemma \ref{lem:onscom}
we find
$
a_m^{\downarrow \Downarrow}=-a_m
$
for $m \in \Z$
and 
$
g_l^{\downarrow \Downarrow}=g_l
$
for $l \in \N$.
We now show that
$a_0^{\prime \downarrow \Downarrow}= -a_1'$. 
We mentioned earlier
that $a_0'=X_{23}$
and 
 $a_1=X_{03}$.
From the former  and
(\ref{eq:downarrow})
we find
$a_0^{\prime\downarrow \Downarrow}= X_{10}$.
From the latter and
(\ref{def:primemap})
we get
$a_1'=X_{01}$.
By these remarks and
since 
$X_{10}=-X_{01}$ we find
$a_0^{\prime \downarrow \Downarrow}= -a_1'$. 
Applying  the automorphism 
$\downarrow \Downarrow$ to
(\ref{eq:act1})--(\ref{eq:act3})
using the above comments
we obtain
(\ref{eq:act4})--(\ref{eq:act6}).
\hfill $\Box $  

\begin{lemma}
\label{lem:3subalg}
Each of the following is a subalgebra of
$\boxtimes$:
\begin{eqnarray*}
\Omega+\Omega',
\qquad \qquad 
\Omega'+\Omega'',
\qquad \qquad 
\Omega+\Omega''.
\end{eqnarray*}
\end{lemma}
\noindent {\it Proof:} 
We first show 
that $\Omega+\Omega'$ is a subalgebra of $\boxtimes$.
Since each of $\Omega, \Omega'$ is a subalgebra
of $\boxtimes$ it suffices to show
 $\lbrack \Omega, \Omega'\rbrack \subseteq 
\Omega+\Omega'$.
Using Lemma
\ref{lem:action}
we find that $\Omega+\Omega'$ is
an invariant subspace for $\mbox{ad}(a_0')$
and 
$\mbox{ad}(a_1')$.
Observe $a_0', a_1'$ generate $\Omega'$
so 
$\Omega+\Omega'$ is an invariant subspace for
$\mbox{ad}(\Omega')$. It follows
that 
 $\lbrack \Omega, \Omega'\rbrack \subseteq 
\Omega+\Omega'$ 
so 
$\Omega+\Omega'$ 
is
a subalgebra of 
$\boxtimes$.
Repeatedly applying the automorphism $\prime$
we find that each of
$\Omega'+\Omega''$,
$\Omega+\Omega''$ is a subalgebra of
$\boxtimes$.
\hfill $\Box $  

\begin{proposition} 
\label{prop:boxspan}
The following (i), (ii) hold:
\begin{enumerate}
\item
 $\boxtimes= \Omega + \Omega'+\Omega''$.
\item
$\boxtimes$ is spanned by the union of
(\ref{eq:omegaspan})--(\ref{eq:omegaprimeprimespan}).
\end{enumerate}
\end{proposition}
\noindent {\it Proof:} 
(i) 
By Lemma
\ref{lem:3subalg}
and since each of 
 $\Omega,\Omega',\Omega''$ is a subalgebra of $\boxtimes$
we find
 $\Omega + \Omega'+\Omega''$ is a subalgebra of $\boxtimes$.
This subalgebra contains the generators 
$\lbrace X_{ij}\;|\;i,j \in \I, i\not=j\rbrace$  for $\boxtimes$
by Definition
\ref{def:omega} and
Lemma \ref{lem:omegap}. Therefore
 $\boxtimes =\Omega + \Omega'+\Omega''$.
\\
\noindent (ii)
Combine (i) above with Lemma
\ref{lem:fourpart}.
\hfill $\Box $

\begin{note}
\rm In Section 11 we will show that
the sum
$\boxtimes = \Omega+\Omega'+\Omega''$ is direct,
and that
the union of 
(\ref{eq:omegaspan})--(\ref{eq:omegaprimeprimespan})
is a basis for $\boxtimes$.
\end{note}

\section{Comments on
$L(\mathfrak{sl}_2)^+$}

\noindent
In this section we shift our attention
to $L(\mathfrak{sl}_2)^+$.
We will define a subalgebra
$\Delta$ of
$L(\mathfrak{sl}_2)^+$ and prove
\begin{eqnarray*}
L(\mathfrak{sl}_2)^+ = \Delta + \Delta' + \Delta''
\qquad \qquad  (\mbox{\rm direct sum}).
\end{eqnarray*}
Later in the paper it
 will turn out that 
$\Delta$ is the image of $\Omega$ under
the map $\sigma$
from
Proposition
\ref{prop:isostart}.

\medskip
\noindent Before proceeding we sharpen
our notation. Referring to
Definition
\ref{def:extsl2}
and
Lemma
\ref{lem:order3},
for $S \in \lbrace T, T',T''\rbrace $ we identify
$\K\lbrack S \rbrack$ with the subalgebra
of $\mathcal A$ generated by $S$.

\begin{definition}
\rm
\label{def:delta}
We let $\Delta$ denote the following subspace of 
$L(\mathfrak{sl}_2)^+$:
\begin{eqnarray}
\label{eq:deltadef}
\Delta = X\otimes \K\lbrack T \rbrack
+ 
 Y\otimes T\K \lbrack T \rbrack
+
 Z\otimes (T-1)\K \lbrack T \rbrack.
\end{eqnarray}
\end{definition}

\begin{lemma}
We have
\begin{eqnarray*}
\Delta' &=& 
 X\otimes (T'-1)\K \lbrack T' \rbrack
+
Y\otimes \K\lbrack T' \rbrack
+ 
 Z\otimes T'\K \lbrack T' \rbrack,
\\
\Delta'' &=&
 X\otimes T''\K \lbrack T'' \rbrack
+
 Y\otimes (T''-1)\K \lbrack T'' \rbrack
+
Z\otimes \K\lbrack T'' \rbrack.
\end{eqnarray*}
\end{lemma}
\noindent {\it Proof:} 
Routine using
Lemma 
\ref{def:prime2}
and Lemma
\ref{lem:sl2aut}(i).
\hfill $\Box $

\begin{lemma}
\label{lem:subalg}
Each of $\Delta, \Delta', \Delta''$ is a subalgebra of
$L(\mathfrak{sl}_2)^+$.
\end{lemma}
\noindent {\it Proof:} 
Using 
(\ref{eq:equit}) and 
(\ref{eq:brackdef}) we find
$\Delta$ is closed under the Lie bracket.
Therefore $\Delta$ is a subalgebra of 
$L(\mathfrak{sl}_2)^+$.
By this and since the map $\prime$ is an automorphism
we find
each of 
$\Delta', \Delta''$ is a subalgebra of 
$L(\mathfrak{sl}_2)^+$.
\hfill $\Box $

\begin{proposition}
\label{prop:deltadec}
We have
\begin{eqnarray*}
L(\mathfrak{sl}_2)^+ = \Delta + \Delta' + \Delta''
\qquad \qquad  (\mbox{\rm direct sum}).
\end{eqnarray*}
\end{proposition}
\noindent {\it Proof:} 
The elements $X,Y,Z$ form a basis for
$\mathfrak{sl}_2$
so
\begin{eqnarray*}
L(\mathfrak{sl}_2)^+ =
X \otimes {\mathcal A}
+ 
Y \otimes {\mathcal A}
+
Z \otimes {\mathcal A}
\qquad \qquad  (\mbox{direct sum}).
\end{eqnarray*}
From Lemma
\ref{lem:abasis} we find
\begin{eqnarray*}
{\mathcal A} = 
 \K\lbrack T \rbrack
+ 
 (T'-1)\K \lbrack T' \rbrack
+
 T''\K \lbrack T'' \rbrack
\qquad \qquad  (\mbox{direct sum})
\end{eqnarray*}
and this implies
\begin{eqnarray}
X\otimes {\mathcal A} = 
X\otimes \K\lbrack T \rbrack
+ 
X\otimes (T'-1)\K \lbrack T' \rbrack
+
X\otimes T''\K \lbrack T'' \rbrack
\qquad \quad  (\mbox{direct sum}).
\label{eq:line1}
\end{eqnarray}
We apply the automorphism
$\prime $
to
(\ref{eq:line1}) and in the resulting equation
 cyclically permute the
terms on the right-hand side. This gives
\begin{eqnarray}
Y\otimes {\mathcal A} &=& 
Y\otimes T\K \lbrack T \rbrack
+
Y\otimes \K\lbrack T' \rbrack
+ 
Y\otimes (T''-1)\K \lbrack T'' \rbrack
\qquad \quad  (\mbox{direct sum}).
\label{eq:line2}
\end{eqnarray}
We apply the automorphism
$\prime $
to
(\ref{eq:line2}) and in the resulting equation
 cyclically permute the
terms on the right-hand side. This gives
\begin{eqnarray}
Z\otimes {\mathcal A} &=& 
Z\otimes (T-1)\K \lbrack T \rbrack
+
Z\otimes T'\K \lbrack T' \rbrack
+
Z\otimes \K\lbrack T'' \rbrack
\qquad \quad  (\mbox{direct sum}).
\label{eq:line3}
\end{eqnarray}
Using Definition
\ref{def:delta} we find
$\Delta$ is the sum of
the first terms on the right in
lines (\ref{eq:line1}),
 (\ref{eq:line2}),
 (\ref{eq:line3}).
Similarly $\Delta'$ (resp. $\Delta''$) is the sum of
the second terms (resp. third terms) 
on the right in
lines (\ref{eq:line1}),
 (\ref{eq:line2}),
 (\ref{eq:line3}).
The result follows.
\hfill $\Box $

\section{Some polynomials}

\noindent In this section we recall the Chebyshev
 polynomials. We will use the following notation.
Let $\lambda $ denote an indeterminate.
Let $\K\lbrack \lambda \rbrack $ denote
the $\K$-algebra consisting of all polynomials
in $\lambda$ that have coefficients in $\K$.

\begin{definition}
\rm 
\cite[p.~101]{AAR}
For an integer $n\geq 0$ we let $U_n$ denote the polynomial in
$\K\lbrack \lambda \rbrack$
that satisfies
\begin{eqnarray*}
U_n\biggl(\frac{\lambda+\lambda^{-1}}{2}\biggr) = 
\frac{\lambda^{n+1}-\lambda^{-n-1}}{\lambda - \lambda^{-1}}.
\end{eqnarray*}
We call $U_n$ the 
{\it nth Chebyshev polynomial of the second kind}.
\end{definition}

\begin{example}
\rm
We have
\begin{eqnarray*}
&&U_0 =1,
\qquad \qquad U_1 = 2\lambda,
\qquad \qquad U_2 = 4\lambda^2-1,
\qquad \qquad
U_3 = 8\lambda^3-4\lambda,
\\
&&\qquad \quad U_4 = 16\lambda^4-12\lambda^2+1,
\qquad \qquad U_5 = 32\lambda^5-32\lambda^3+6\lambda.
\end{eqnarray*}
\end{example}

\begin{lemma} 
\label{lem:ttr}
\cite[Section~1.8.2]{KoeSwa}
The Chebyshev polynomials satisfy the following 3-term
recurrence:
\begin{eqnarray*}
&&2\lambda U_n =  U_{n+1}+U_{n-1} \qquad \qquad n=0,1,\ldots
\\
&& \qquad U_0=1, \qquad \qquad U_{-1} = 0.
\end{eqnarray*}
\end{lemma}

\begin{lemma} 
\cite[Section~1.8.2]{KoeSwa}
The Chebyshev polynomials have the following
presentation in terms of hypergeometric series:
\begin{eqnarray*}
U_n(\lambda) = (n+1)\; 
 {{}_2}F_1\Biggl({{-n,\, n+2}\atop {3/2}}
\;\Bigg\vert \;\frac{1-\lambda}{2}\Biggr)
\qquad \qquad n = 0,1,2,\ldots
\end{eqnarray*}
\end{lemma}

\noindent We have a comment.

\begin{lemma}
\label{lem:polybasis}
The following is a basis for the
$\K$-vector space 
$\K\lbrack \lambda \rbrack$:
\begin{eqnarray*}
U_n(1-2\lambda)   \qquad \qquad n=0,1,2,\ldots   
\end{eqnarray*}
\end{lemma}
\noindent {\it Proof:} 
For an integer $n\geq 0$ the polynomial
$U_n$ has degree exactly $n$ by
Lemma
\ref{lem:ttr}. From this we find
$U_n(1-2\lambda)$ has degree exactly
$n$ as a polynomial in $\lambda$.
The result follows.
\hfill $\Box $

\section{A basis for 
 $L(\mathfrak{sl}_2)^+$}

\noindent In 
Section 8 we obtained a direct sum
decompostion 
 $L(\mathfrak{sl}_2)^+=\Delta+\Delta'+\Delta''$.
In this section we find a basis for each of
 $\Delta, \Delta',\Delta''$. The union of these
 bases is a basis
 for 
 $L(\mathfrak{sl}_2)^+$ that we will find useful later in the paper.

\begin{lemma}
\label{lem:bpart1}
Referring to Definition
\ref{def:delta}
the following (i)--(iv) hold.
\begin{enumerate}
\item $X\otimes \K\lbrack T \rbrack$ has a basis
\begin{eqnarray}
\label{eq:b1}
X\otimes U_{m-1}(1-2T) \qquad \qquad m \in \N.
\end{eqnarray}
\item $Y\otimes T\K\lbrack T \rbrack$ has a basis
\begin{eqnarray}
\label{eq:b2}
Y\otimes TU_{m-1}(1-2T) \qquad \qquad m \in \N.
\end{eqnarray}
\item $Z\otimes (T-1)\K\lbrack T \rbrack$ has a basis
\begin{eqnarray}
\label{eq:b3}
Z\otimes (T-1)U_{m-1}(1-2T) \qquad \qquad m \in \N.
\end{eqnarray}
\item The union of 
(\ref{eq:b1})--(\ref{eq:b3}) is a basis for
$\Delta$.
\end{enumerate}
\end{lemma}
\noindent {\it Proof:} 
The assertions (i)--(iii) follow from
Lemma
\ref{lem:polybasis}. Assertion (iv) follows
from (i)--(iii) and
since the sum 
(\ref{eq:deltadef})
is direct.
\hfill $\Box $

\begin{lemma}
\label{lem:bpart2}
The following
(i)--(iv) hold.
\begin{enumerate}
\item $X\otimes (T'-1)\K\lbrack T' \rbrack$ has a basis
\begin{eqnarray}
\label{eq:b32}
X\otimes (T'-1)U_{m-1}(1-2T') \qquad \qquad m \in \N.
\end{eqnarray}
\item 
$Y\otimes \K\lbrack T' \rbrack$ has a basis
\begin{eqnarray}
\label{eq:b12}
Y\otimes U_{m-1}(1-2T') \qquad \qquad m \in \N.
\end{eqnarray}
\item
$Z\otimes T'\K\lbrack T' \rbrack$ has a basis 
\begin{eqnarray}
\label{eq:b22}
Z\otimes T'U_{m-1}(1-2T') \qquad \qquad m \in \N.
\end{eqnarray}
\item The union of 
(\ref{eq:b32})--(\ref{eq:b22}) is a basis for
$\Delta'$.
\end{enumerate}
\end{lemma}
\noindent {\it Proof:} 
Apply the automorphism $\prime $ to
the vectors in 
Lemma
\ref{lem:bpart1}.
\hfill $\Box $

\begin{lemma}
\label{lem:bpart3}
The following 
(i)--(iv) hold.
\begin{enumerate}
\item  $X\otimes T''\K\lbrack T'' \rbrack$ has a basis
\begin{eqnarray}
\label{eq:b23}
X\otimes T''U_{m-1}(1-2T'') \qquad \qquad m \in \N.
\end{eqnarray}
\item $Y\otimes (T''-1)\K\lbrack T'' \rbrack$ has a basis 
\begin{eqnarray}
\label{eq:b33}
Y\otimes (T''-1)U_{m-1}(1-2T'') \qquad \qquad m \in \N.
\end{eqnarray}
\item $Z\otimes \K\lbrack T'' \rbrack$ has a basis 
\begin{eqnarray}
\label{eq:b13}
Z\otimes U_{m-1}(1-2T'') \qquad \qquad m \in \N.
\end{eqnarray}
\item The union of 
(\ref{eq:b23})--(\ref{eq:b13}) is a basis for
$\Delta''$.
\end{enumerate}
\end{lemma}
\noindent {\it Proof:} 
Apply the automorphism $\prime $ to
the vectors in 
Lemma
\ref{lem:bpart2}.
\hfill $\Box $

\begin{theorem}
\label{thm:9basis}
The union of 
(\ref{eq:b1})--(\ref{eq:b13}) is a basis for
the $\K$-vector space $L(\mathfrak{sl}_2)^+$.
\end{theorem}
\noindent {\it Proof:} 
Combine
Proposition
\ref{prop:deltadec}
with 
Lemma
\ref{lem:bpart1}(iv),
Lemma
\ref{lem:bpart2}(iv),
Lemma
\ref{lem:bpart3}(iv).
\hfill $\Box $

\section{The main results
}

In this section we show that  
the sum
$\boxtimes = \Omega+\Omega'+\Omega''$ is direct,
and that
the union of 
(\ref{eq:omegaspan})--(\ref{eq:omegaprimeprimespan})
is a basis for $\boxtimes$.
We also show that the Lie algebra homomorphism
$\sigma :\boxtimes \rightarrow
L(\mathfrak{sl}_2)^+$ from
Proposition
\ref{prop:isostart}
is an isomorphism.
Our proofs are based on the
following proposition, in which
 we apply 
$\sigma $ to
(\ref{eq:omegaspan})--(\ref{eq:omegaprimeprimespan})
and express the image in terms of the basis
from Theorem \ref{thm:9basis}.

\begin{proposition}
\label{thm:sigimage}
Let the homomorphism $\sigma: \boxtimes \to 
L(\mathfrak{sl}_2)^+$  be as in
Proposition
\ref{prop:isostart}. 
Then for each element
$u$ in the table below, the expression to
the right of $u$ is the image of $u$ under
$\sigma$.
\bigskip

\centerline{
\begin{tabular}[t]{c|c}
{\rm element $u$} & {\rm image of $u$ under $\sigma$}
\\ \hline  \hline
\\
$a_m$
&
$
-X\otimes U_{m-2}(1-2T)
+
Y\otimes
TU_{m-1}(1-2T)
+
Z\otimes 
(T-1)U_{m-1}(1-2T)
$
\\
$a_{1-m}$
&
$
X\otimes U_{m-1}(1-2T)
-
Y\otimes
TU_{m-2}(1-2T)
-
Z\otimes 
(T-1)U_{m-2}(1-2T)
$
\\
$g_m$
&
$
-X\otimes U_{m-1}(1-2T)
-
Y\otimes
TU_{m-1}(1-2T)
+
Z\otimes 
(T-1)U_{m-1}(1-2T)
$
\\
\\
\hline
\\
$a'_m$
&
$
X\otimes 
(T'-1)U_{m-1}(1-2T')
-Y\otimes U_{m-2}(1-2T')
+
Z\otimes
T'U_{m-1}(1-2T')
$
\\
$a'_{1-m}$
&
$
-X\otimes 
(T'-1)U_{m-2}(1-2T')
+
Y\otimes U_{m-1}(1-2T')
-
Z\otimes
T'U_{m-2}(1-2T')
$
\\
$g'_m$
&
$
X\otimes 
(T'-1)U_{m-1}(1-2T')
-
Y\otimes U_{m-1}(1-2T')
-
Z\otimes
T'U_{m-1}(1-2T')
$
\\
\\
\hline
\\
$a''_m$
&
$
X\otimes
T''U_{m-1}(1-2T'')
+
Y\otimes 
(T''-1)U_{m-1}(1-2T'')
-Z\otimes U_{m-2}(1-2T'')
$
\\
$a''_{1-m}$
&
$
-X\otimes
T''U_{m-2}(1-2T'')
-
Y\otimes 
(T''-1)U_{m-2}(1-2T'')
+
Z\otimes U_{m-1}(1-2T'')
$
\\
$g''_m$
&
$
-
X\otimes
T''U_{m-1}(1-2T'')
+
Y\otimes 
(T''-1)U_{m-1}(1-2T'')
-
Z\otimes U_{m-1}(1-2T'')
$
\end{tabular}
}
\bigskip
\noindent
In the above table we assume $m \in \N$.
\end{proposition}
\noindent {\it Proof:} 
Referring to the above table, 
for $m \in \N$ 
let ${\hat a}_m$, ${\hat a}_{1-m}$,
and 
${\hat g}_m$
denote
the expressions to the right of
${a}_m$, ${a}_{1-m}$, and
${g}_m$ respectively.
We show
 $a_m^\sigma
 ={\hat a}_m$,
$a_{1-m}^\sigma=
 {\hat a}_{1-m}$,
and 
${g}_m^\sigma
={\hat g}_m$.
Using the data in the above table
we find
\begin{eqnarray*}
{\hat a}_0 = X\otimes 1,
\qquad \qquad 
{\hat a}_1 = Y\otimes T +Z\otimes(T-1).
\end{eqnarray*}
Using the data in the above table
and
(\ref{eq:equit}), Lemma
\ref{lem:ttr} we find
\begin{eqnarray*}
\lbrack {\hat a}_m, {\hat a}_0 \rbrack
&=& 2 {\hat g}_m,
\\
\lbrack {\hat g}_1, {\hat a}_m \rbrack &=& 
{\hat a}_{m+1}-{\hat a}_{m-1},
\\
\lbrack {\hat g}_1, {\hat a}_{1-m} \rbrack &=& 
{\hat a}_{2-m}-{\hat a}_{-m}
\end{eqnarray*}
for $m \in \N$.
Recall by Lemma
\ref{lem:a0a1} that
$a_0=X_{12}$,
$a_1=X_{03}$. By this and 
Proposition
\ref{prop:isostart} we find
\begin{eqnarray*}
a_0^\sigma = X\otimes 1,
\qquad \qquad 
a_1^\sigma = Y\otimes T +Z\otimes(T-1).
\end{eqnarray*}
By 
Lemma
\ref{lem:a0a1}
and since $\sigma$ is a homomorphism of
Lie algeras,
\begin{eqnarray*}
\lbrack a_m^\sigma, a_0^\sigma \rbrack
&=& 2 g_m^\sigma,
\\
\lbrack g_1^\sigma, a_m^\sigma \rbrack 
&=& a_{m+1}^\sigma-a_{m-1}^\sigma,
\\
\lbrack {g}_1^\sigma, a_{1-m}^\sigma \rbrack &=& 
a_{2-m}^\sigma-a_{-m}^\sigma
\end{eqnarray*}
for $m \in \N$.
By these comments the  
 ${\hat a}_m, {\hat a}_{1-m},
{\hat g}_m$
and the
 $a_m^\sigma,
a_{1-m}^\sigma,
{g}_m^\sigma$ satisfy the
same recursion and the same initial conditions.
It follows
 $a_m^\sigma
 ={\hat a}_m$,
$a_{1-m}^\sigma=
 {\hat a}_{1-m}$,
and 
${g}_m^\sigma
={\hat g}_m$ for $m \in \N$.
We have now verified the upper third of the
table.  To verify the
remaining two thirds of the table,
apply the automorphism $\prime$
and 
use
Lemma
\ref{lem:comdiag}.
\hfill $\Box $  

\begin{lemma}
\label{lem:basisdel}
The following (i)--(iv) hold.
\begin{enumerate}
\item
$\Delta$ has a basis 
\begin{eqnarray}
\label{eq:basisdel1}
a_m^\sigma, \;
g_l^\sigma \qquad \qquad m\in \Z, \quad l\in \N.
\end{eqnarray}
\item
 $\Delta'$ has a basis
\begin{eqnarray}
\label{eq:basisdel2}
a_m^{\prime\sigma}, 
\;g_l^{\prime \sigma} \qquad \qquad m\in \Z, \quad l\in \N.
\end{eqnarray}
\item
$\Delta''$ has a basis
\begin{eqnarray}
\label{eq:basisdel3}
a_m^{\prime \prime \sigma}, \; 
g_l^{\prime \prime \sigma} \qquad \qquad m\in \Z, \quad l\in \N.
\end{eqnarray}
\item The union of 
(\ref{eq:basisdel1})--(\ref{eq:basisdel3})
is a basis for
$L(\mathfrak{sl}_2)^+$. 
\end{enumerate}
\end{lemma}
\noindent {\it Proof:} 
(i)
The elements
(\ref{eq:basisdel1}) are contained in
$\Delta$ by 
Definition
\ref{def:delta} and 
the data in the table of
Proposition
\ref{thm:sigimage}.
In Lemma
\ref{lem:bpart1}(iv) we gave a basis
for $\Delta$. Consider the following ordering
of the vectors in this basis:

\parbox{10cm}{
\begin{eqnarray*}
&&X\otimes 1, \quad  Y\otimes T,  \quad Z\otimes(T-1),
\quad X\otimes U_1(1-2T), 
\nonumber
\\
&&\qquad \qquad \qquad  \qquad 
Y\otimes TU_1(1-2T), \quad Z \otimes (T-1)U_1(1-2T), \quad \ldots
\end{eqnarray*}} \hfill
\parbox{1cm}{\begin{eqnarray}
\label{eq:deltabasislist}
\end{eqnarray}}

\noindent Now consider the sequence
\begin{eqnarray}
a_0^\sigma,\quad  
a_1^\sigma-g_1^\sigma, \quad
g_1^\sigma,\quad
a_{-1}^\sigma, \quad 
a_2^\sigma-g_2^\sigma,\quad
g_2^\sigma,\quad
\ldots
\label{eq:siglist}
\end{eqnarray}
Using the table in
Proposition
\ref{thm:sigimage} we
express each 
vector in
(\ref{eq:siglist})
as a linear combination
of
(\ref{eq:deltabasislist}).
We observe that  the corresponding matrix of coefficients
is upper
triangular
with all diagonal entries nonzero.
By this and since
(\ref{eq:deltabasislist}) is a basis
for $\Delta$ we find
(\ref{eq:siglist}) is a basis for $\Delta$.
From this we routinely find that
(\ref{eq:basisdel1}) is a basis for $\Delta$.
\\
\noindent (ii), (iii) Apply the automorphism
$\prime $
and use
Lemma
\ref{lem:comdiag}.
\\
\noindent (iv) Use Proposition
\ref{prop:deltadec}
and (i)--(iii) above.
\hfill $\Box $  

\begin{corollary}
\label{cor:image}
Under the map
$\sigma : \boxtimes \to 
L(\mathfrak{sl}_2)^+$ 
from
Proposition \ref{prop:isostart},
the image of
$\Omega, \Omega',\Omega''$ is
$\Delta, \Delta',\Delta''$ respectively.
\end{corollary}
\noindent {\it Proof:} 
We first show that $\Delta$ is the image
$\Omega^\sigma$.
The vectors $a_m, g_l,$ $m \in \Z, l \in \N$ span
$\Omega$ by
Lemma
\ref{lem:fourpart}(i). 
Therefore  the vectors
 $a_m^\sigma, g_l^\sigma,$ $m \in \Z, l \in \N$ span
$\Omega^\sigma$.
But these vectors span 
$\Delta$ by
Lemma
\ref{lem:basisdel}(i) so
$\Omega^\sigma=\Delta$.
We have now shown that $\Delta$ is the image of
$\Omega$ under $\sigma$. Our remaining assertions
follow from this and Lemma
\ref{lem:comdiag}. 
\hfill $\Box $  

\begin{theorem}
\label{cor:extbasis}
The following
(i)--(iv) hold.
\begin{enumerate}
\item
The elements 
(\ref{eq:omegaspan})
form a basis for $\Omega$.
\item
The elements 
(\ref{eq:omegaprimespan})
form a basis for $\Omega'$.
\item
The elements 
(\ref{eq:omegaprimeprimespan})
form a basis for $\Omega''$.
\item The union of
(\ref{eq:omegaspan})--(\ref{eq:omegaprimeprimespan})
is a basis for $\boxtimes$.
\end{enumerate}
\end{theorem}
\noindent {\it Proof:} 
(i) The elements
(\ref{eq:omegaspan}) span
$\Omega $ by 
Lemma
\ref{lem:fourpart}(i). The elements
(\ref{eq:omegaspan})
are linearly independent  since
their images under $\sigma$ are linearly
independent by Lemma
\ref{lem:basisdel}(i).
\\
\noindent (ii), (iii) Similar to the proof of (i) above.
\\
\noindent 
(iv)
 The vectors 
(\ref{eq:omegaspan})--(\ref{eq:omegaprimeprimespan})
span $\boxtimes$
by Proposition
\ref{prop:boxspan}.
The vectors 
(\ref{eq:omegaspan})--(\ref{eq:omegaprimeprimespan})
are linearly independent  since
their images under $\sigma$ are linearly
independent by Lemma
\ref{lem:basisdel}(iv).
\hfill $\Box $

\begin{theorem}
The Lie algebra homomorphism
$\sigma : \boxtimes \to 
L(\mathfrak{sl}_2)^+$ 
from
Proposition \ref{prop:isostart} is an isomorphism.
\end{theorem}
\noindent {\it Proof:} 
The map $\sigma$ sends the basis
for $\boxtimes$ given in
Theorem \ref{cor:extbasis}(iv)
to the basis for 
$L(\mathfrak{sl}_2)^+$  given
in 
Lemma \ref{lem:basisdel}(iv). The result follows.
\hfill $\Box $  

\begin{theorem}
\label{thm:dirs}
The sum
$\boxtimes = \Omega+\Omega'+\Omega''$ is direct.
\end{theorem}
\noindent {\it Proof:} 
Immediate from
Theorem \ref{cor:extbasis}.
\hfill $\Box $  

\section{Conclusion}

\noindent In this section we prove
the outstanding assertions from earlier in the paper.

\begin{corollary}
\label{cor:sl2cleanup}
The Lie algebra homomorphism 
$\mathfrak{sl}_2 \rightarrow \boxtimes$ 
given in
Proposition \ref{lem:sl2inj} is an injection.
\end{corollary}
\noindent {\it Proof:} 
Referring to 
Proposition \ref{lem:sl2inj}
and in view of the $S_4$-action on
$\boxtimes$,
 without
loss we may assume $(h,i,j)=(1,2,3)$ 
so that the homomorphism is standard.
The elements $X,Y,Z$ form a basis for 
$\mathfrak{sl}_2$, and their images
under the standard homomorphism
are $X_{12}, X_{23}, X_{31}$ respectively.
It suffices to show
that
$X_{12}, X_{23}, X_{31}$ are linearly independent.
They are linearly independent since
$X_{12}=a_0$, $X_{23}=a_0'$, $X_{31}=a_0''$ 
and since $a_0, a_0', a_0''$ are linearly 
independent
by 
Theorem
\ref{cor:extbasis}(iv).
\hfill $\Box $  

\begin{corollary}
\label{cor:onscleanup}
The Lie algebra homomorphism 
$O\rightarrow \boxtimes$ 
given in
Proposition
\ref{prop:onshom}
is an injection.
\end{corollary}
\noindent {\it Proof:} 
With reference to
Proposition
\ref{prop:onshom} 
and in view of
the $S_4$-action on
$\boxtimes$,
 without loss
we may assume $(h,i,j,k)=(1,2,0,3)$
so that the homomorphism is standard.
The elements
$A_m,G_l,$ $m\in \Z,l\in\N$ form a basis
for $O$ and their images under the standard homomorphism
are 
$a_m,g_l,$ $m\in \Z,l\in\N$.
Therefore it suffices to show that  
$a_m,g_l,$ $m\in \Z,l\in\N$ are linearly
independent. But this is the case
by 
Theorem \ref{cor:extbasis}(i).
\hfill $\Box $

\begin{corollary}
\label{cor:loopcleanup}
The Lie algebra homomorphism 
 $L(\mathfrak{sl}_2) \rightarrow 
 \boxtimes$ 
given in
Proposition
\ref{prop:loopinj}
is an injection.
\end{corollary}
\noindent {\it Proof:} 
Referring to Proposition
\ref{prop:loopinj}
and in view of
the $S_4$-action on
$\boxtimes$,
without loss we may assume
$(h,i,j,k)=(1,2,3,0)$ so that the homomorphism
is standard.
By Definition
\ref{def:nat} the natural
homomorphism 
$L(\mathfrak{sl}_2) \rightarrow  
L(\mathfrak{sl}_2)^+$
is an injection.
By Lemma
\ref{lem:comdiagnat} the natural homomorphism
is the 
composition of the standard homomorphism
 $L(\mathfrak{sl}_2) \rightarrow 
 \boxtimes$  
and the isomorphism
$\sigma :\boxtimes
\rightarrow 
 L(\mathfrak{sl}_2)^+$.
Therefore 
the standard homomorphism 
 $L(\mathfrak{sl}_2) \rightarrow 
 \boxtimes$  
is an injection. The result follows.
\hfill $\Box $ \\ 

\begin{corollary}
For mutually distinct $h,i,j \in \I$
the elements 
 $X_{hi}, X_{ij}, X_{jh}$ form a basis
 for a subalgebra of 
$\boxtimes$ that is isomorphic to
$\mathfrak{sl}_2$.
\end{corollary} 
\noindent {\it Proof:} 
Immediate from
Proposition \ref{lem:sl2inj}
and Corollary
\ref{cor:sl2cleanup}.
\hfill $\Box $

\begin{corollary}
For mutually distinct 
$h,i,j,k \in \I$
the subalgebra of $\boxtimes$ generated
by $X_{hi}, X_{jk}$ is isomorphic
to the Onsager algebra.
\end{corollary}
\noindent {\it Proof:} 
Immediate from
Proposition
\ref{prop:onshom}
and
Corollary
\ref{cor:onscleanup}.
\hfill $\Box $

\begin{corollary}
For distinct $r,s \in \I$
the subalgebra
of $\boxtimes$ generated by
\begin{eqnarray*}
\lbrace X_{ij} \,|\,i,j\in \I , i\not=j, (i,j)\not=(r,s), (i,j)\not=(s,r)
\rbrace
\end{eqnarray*}
is isomorphic to the loop algebra
 $L(\mathfrak{sl}_2)$.
\end{corollary}
\noindent {\it Proof:} 
This subalgebra is the image of 
 $L(\mathfrak{sl}_2)$ under
 the homomorphism in 
Proposition
\ref{prop:loopinj}, where in that proposition
we take 
$i=r$ and $k=s$.
The result follows in view of Corollary
\ref{cor:loopcleanup}.
\hfill $\Box $   \\

\noindent We finish this section with a comment.

\begin{corollary}
\label{eq:s4inj}
The group homomorphism
$S_4 \rightarrow \mbox{Aut}(\boxtimes)$ from Section 2
is an injection.
\end{corollary}
\noindent {\it Proof:} 
The elements 
\begin{eqnarray*}
X_{12}, X_{23}, X_{31}, X_{03}, X_{01}, X_{02}
\end{eqnarray*}
are linearly independent
since they are the images under the isomorphism
$\sigma $ of
$a_0, a'_0, a''_0$, $a_1, a'_1, a''_1$.
Combining this with
Definition
\ref{def:tet}(i) we find
$\lbrace X_{ij} \,|\,i,j\in \I, i\not=j\rbrace $
are mutually distinct. 
For $\tau$ in the kernel of the
group homomorphism
$S_4 \rightarrow \mbox{Aut}(\boxtimes)$
and for distinct $i,j \in \I$ we
find $X_{ij}=X_{i^\tau j^\tau}$
by
(\ref{eq:perm}), so $i^\tau=i$ and $j^\tau=j$ by our preliminary
remark.
Apparently $\tau$ stabilizes each element of $\I$ so
$\tau$ is the identity element.
The result follows.
\hfill $\Box $  

\section{Suggestions for further research}

In this section we give some suggestions for further research.

\begin{problem}
\label{prob:1}
\rm
By Theorem
\ref{cor:extbasis}
the union of
(\ref{eq:omegaspan})--(\ref{eq:omegaprimeprimespan})
is a basis for the $\K$-vector space $\boxtimes$.
Compute the action of the Lie bracket
on this basis.
See 
Lemma \ref{lem:a0a1}
and
Lemma \ref{lem:action}
for partial results.
\end{problem}

\begin{problem}
\label{prob:2}
\rm
Compute the group
$\mbox{Aut}(\boxtimes)$.
We recall by Corollary
\ref{eq:s4inj}
that 
the homomorphism of groups
$S_4 \rightarrow \mbox{Aut}(\boxtimes)$
given in Section 2 is an injection.
\end{problem}

\begin{problem}
\label{prob:3}
\rm
Find all the ideals in the Lie algebra $\boxtimes$.
\end{problem}

\noindent The following problem was inspired by
\cite{elduque}.

\begin{problem}
\label{prob:4}
\rm
For $i,j \in \lbrace 0,1\rbrace$ we define 
\begin{eqnarray*}
\boxtimes_{ij} =
\lbrace v \in \boxtimes \;|\;v^d=(-1)^iv, \;\;v^*=(-1)^jv\rbrace
\end{eqnarray*}
where the automorphisms  
$d,*$ are from
(\ref{eq:primelab}),
(\ref{eq:downarrow}) respectively.
Since $d,*$ are commuting involutions we find
\begin{eqnarray*}
\boxtimes = 
\boxtimes_{00}
+
\boxtimes_{01}
+
\boxtimes_{10}
+
\boxtimes_{11}
\qquad \qquad (\mbox{direct sum}).
\end{eqnarray*}
By the construction
\begin{eqnarray}
\label{eq:lie}
\lbrack \boxtimes_{ij}, \boxtimes_{rs} \rbrack
\subseteq \boxtimes_{i+r,j+s}
\end{eqnarray}
where the subscripts are computed modulo 2.
Show that $\boxtimes_{00}=0$,
and conclude using
(\ref{eq:lie})
that each of
$\boxtimes_{01},
\boxtimes_{10},
\boxtimes_{11}$ is an abelian subalgebra of $\boxtimes$.
Find a basis for each of these subalgebras.
Investigate the relationship between the decomposition
$\boxtimes = 
\boxtimes_{01}
+\boxtimes_{10}
+
\boxtimes_{11}$ 
and the decomposition
$\boxtimes = \Omega+\Omega'+\Omega''$ from
Theorem
\ref{thm:dirs}.
\end{problem}

\noindent Given the results of this paper it is
natural to consider the following generalization
of the algebra $\boxtimes$.

\begin{problem}
\label{prob:5}
\rm
By a {\it graph} 
we mean a pair
$\Gamma=(X,E)$ where
$X$ is a nonempty finite set 
and $E \subseteq X^2$ is a binary
relation such that 
$ii\not\in E$ for all $i \in X$ and
$ij \in E \Leftrightarrow ji\in E$ for all $i,j\in X$.
Given a graph $\Gamma=(X,E)$
let $\mathcal L={\mathcal L}(\Gamma)$ denote the Lie algebra over $\K$ 
that has generators
\begin{eqnarray*}
\lbrace X_{ij} \,|\,i,j\in X, \quad ij \in E\rbrace
\end{eqnarray*}
and the following relations:
\begin{enumerate}
\item For  $ij\in E$,
\begin{eqnarray*}
X_{ij}+X_{ji} = 0.
\end{eqnarray*}
\item For $hi \in E$ and $ij \in E$,
\begin{eqnarray*}
\lbrack X_{hi},X_{ij}\rbrack = 2X_{hi}+2X_{ij}.
\end{eqnarray*}
\item For  $hi \in E$ and $jk \in E$,
\begin{eqnarray*}
\lbrack X_{hi},
\lbrack X_{hi},
\lbrack X_{hi},
X_{jk}\rbrack \rbrack \rbrack= 
4 \lbrack X_{hi},
X_{jk}\rbrack.
\end{eqnarray*}
\end{enumerate}
Given a subset $E_1 \subseteq E$
the inclusion map $E_1\to E$ induces
a homomorphism of Lie algebras
${\mathcal L}(\Gamma_1) \to 
{\mathcal L}(\Gamma)$, where
$\Gamma_1=(X,E_1)$. 
Show that this homomorphism is an injection.
\end{problem}

\noindent Brian Hartwig \hfil\break
\noindent Department of Mathematics \hfil\break
\noindent University of Wisconsin \hfil\break
\noindent 480 Lincoln Drive \hfil\break
\noindent Madison, WI 53706-1388 USA \hfil\break
\noindent email: {\tt hartwig@math.wisc.edu }\hfil\break

\bigskip

\noindent Paul Terwilliger \hfil\break
\noindent Department of Mathematics \hfil\break
\noindent University of Wisconsin \hfil\break
\noindent 480 Lincoln Drive \hfil\break
\noindent Madison, WI 53706-1388 USA \hfil\break
\noindent email: {\tt terwilli@math.wisc.edu }\hfil\break

\end{document}